\documentstyle[aps,prl,twocolumn]{revtex}

\begin{document}
\draft

\twocolumn[\hsize\textwidth\columnwidth\hsize\csname@twocolumnfalse\endcsname
\title{Possible Localized Modes in the Uniform Quantum Heisenberg Chains of Sr$_{2}$CuO$_{3}$.}

\author{J.P. Boucher$^{a,b}$ and M. Takigawa$^{c}$}

\address{$^{a}$ Laboratoire de Spectrom\'{e}trie Physique, Universit\'{e} J. Fourier
Grenoble I, BP 87, 38402 Saint Martin d'H\`{e}res cedex, France.\\
$^{b}$ Department of Physics, Kyushu University 33, Fukuoka, 812-8581, Japan.\\
$^{c}$ Institute for Solid State Physics, University of Tokyo, Roppongi,
Minato-ku, Tokyo 106, Japan.}
\date{25 Septembre 1999}

\maketitle

\begin{abstract}
A model of mobile bond-defects is tentatively proposed to analyze the
``anomalies'' observed on the NMR spectrum of the quantum Heisenberg chains
of Sr$_{2}$CuO$_{3}$. A bond-defect is a local change in the exchange coupling. It
results in a local alternating magnetization (LAM), which, when the defect
moves, creates a flipping proces of the local field seen by each nuclear
spin. At low temperature, when the overlap of the LAM becomes large,
the defects form a periodic structure, which extends over almost all the chains. 
In that regime, the density of bond-defects decreases linearly with
T.
\end{abstract}


\

] \narrowtext

In low-dimensional quantum antiferromagnets, the effects of impurities or
defects in ``gapped'' and ``ungapped'' systems may display drastic
differences but also common features\cite{1Laukamp98}. A good example of the
latter case is provided by the uniform quantum ($s=1/2$) Heisenberg chains
(UQHC). Hereafter, we consider the effects due to defects in such systems. A
``defect'' is defined as a local change in the magnetic bond coupling. As
the translational symmetry of the spin system is broken, a local alternating
magnetization (LAM) develops around the defects in the direction of the
applied field: this is one of the common features mentioned above\cite
{1Laukamp98}. Such a LAM is probed accurately by NMR\cite{2Takigawa97}\cite
{3Fagot96}. Our analysis relies on results obtained on the compound Sr$_{2}$%
CuO$_{3}$\cite{2Takigawa97}. In such a ``real system'', the elastic coupling
between the spins and the lattice cannot be ignored. It can result in a
lattice distortion giving rise to the so-called spin-Peierls transition (at
temperature $T_{sp}$)\cite{4Bray83}. Dynamical effects on LAM structures can
also been expected from such magneto-elastic couplings. In the dimerized (D)
phase of a spin-Peierls system (SPS), i.e., for $T<T_{sp}$, the LAM induced
by chain-end effects become mobile in the pure one-dimensional case\cite
{5Hansen98}. In the ``high-field'' phase of a SPS, a periodic LAM structure
occurs, which results from the field-induced incommensurate lattice
modulation\cite{6Uhrig98}. In that case, dynamical effects are due to the
quantum fluctuations of the phason modes describing the vibrations of the
modulated lattice structure\cite{6Uhrig98}. In the uniform phase of a SPS,
i.e. for $T>T_{sp}$, the spin-lattice coupling is also of crucial
importance. It can appreciably change the magnetic susceptibility of the
system well above $T_{sp}$\cite{7Sandvik97}.

In Sr$_{2}$CuO$_{3}$, no spin-Peierls transition is observed as a
3-dimensional magnetic ordering takes place at $T_{N}\simeq 5$ $K$. Above $%
T_{N}$, this compound provides a remarkable example of UQHC. The exchange
coupling is large ($J\simeq 2200$ K). The logarithmic corrections
characterizing the quantum ($T<<J$) behavior of UQHC have been observed\cite
{8Motoyama}. LAM associated with the edges of finite chains have also been
observed in Sr$_{2}$CuO$_{3}$\cite{2Takigawa97}. Due to that LAM, the NMR
lineshape is changed in a very specific way, giving rise to ``features'',
which have been well identified. This is the case of features ``A''
displayed in Figs. 1 and 2. Other features, however, are visible on the NMR
line (features ``B'' and ``C'' in Figs. 1 and 2), which have not been
explained yet. We propose an analysis of these additional features in terms
of LAM associated with mobile ``bond-defects''.

The NMR spectra to be analyzed - a few examples are displayed in Figs. 1 and
2 - have been obtained on an Ar-annealed single crystal. That annealing
procedure minimizes in that compound the possible source of impurities
induced by interstitial excess oxygen. After such a treatment, the
concentration of the residual spin-1/2 impurities becomes as low as $1.3$ $%
10^{-4}$ \cite{8Motoyama}. Some of the NMR spectra displayed in Figs. 1 and
2 have been previously reported (Figs. 1a, b and Fig. 2a) in\cite
{2Takigawa97} together with the experimental conditions. As clearly
established in\cite{2Takigawa97}, features ``A'' originate from the chain
edges, while we show here that features ``B'' is well explained by the
presence of mobile bond-defects. Features ``C'' would result from the
periodic structure they form at low $T$. Finally, possible origins of these
defects are suggested in relation with the lattice properties.

First we consider features ``A''. LAM associated with the edges of finite
UQHC have been described by Eggert and Affleck\cite{9Eggert92}. The LAM
amplitude increases with distance from the chain end. At $T\neq 0$ however,
thermal fluctuations act as a cutoff, and at long distance, the LAM
amplitude decreases exponentially with a characteristic scale given by the
correlation length of the system: $\xi =J/2T$. An analytical expression$\ $%
for the local susceptibility has been derived, which agrees well with
Monte-Carlo calculations. Accordingly, for an hyperfine coupling $A_{\beta }$
(expressed in GHz and $\beta $ = {\bf a}, {\bf b} or {\bf c} referring to
the field direction with respect to the crystal axes - chain axis {\bf b}),
the local field seen by a nuclear spin at site n (counted from a chain end)
can be written 
\[
h_{n}=-(-1)^{n}B\text{ }A_{\beta }\text{ }H\text{ }\{(n+\phi )/\sqrt{\xi
\sinh (2(n+\phi )/\xi }~~(1) 
\]
with $B\simeq 0.020$ GHz$^{-1}$ and $h_{n}$ and $H$ expressed in Tesla. In
(1), $\phi $ allows for a small shift of the LAM along the chain.$~\phi $
changes the position $n_{max}$(see Fig. 3a) of the maximum field value, but
for $\phi \leq $\ $n_{max}$, it does not affect this value:$~h_{max}\simeq
1.08~10^{-2}~A_{\beta }~H\sqrt{\xi }$ ($A_{a}\simeq 0.14$ GHz and $%
A_{c}\simeq 0.50$ GHz). In the NMR spectrum, the field splitting $\Delta H$
(see Fig. 1) where features ``A'' occur is given by $\Delta H=2h_{max}$. In
that respect, as explained in \cite{2Takigawa97}, a very good agreement is
obtained between theory and experiment. The shape of features ``A'',
however, depends appreciably on $\phi $ (see dot line $\phi =0$ in Fig. 1b).
Using $\phi $ as an adjustable parameter, a good agreement with the
experimental line shape can be achieved for all the collected data (see dots
in Figs. 1 and 2). With that procedure, one obtains that $\phi $ is field
independent. It shows, however, a $T$ dependence: $\phi =(1200\pm 200)/T$,
which, remarkably, agrees with $\phi \simeq \xi $. As shown in Fig. 3a
(solid line), for that particular value of $\phi $, the LAM displays no
maximum, but a rather ``flat'' initial behavior. Despite the rough
description used here, one is led to conclude that the initial increase of
the LAM amplitude is not observed experimentally, in contradiction with the
theoretical prediction\cite{9Eggert92}. The elastic spin-lattice coupling
may play here an important role. The model of a sudden cutoff of the
exchange coupling (J = 0) just at the chain edges would not applied to such
``real'' systems. Instead, as in the D phase of a SPS\cite{10Augier98}, on a
length scale $\simeq $ $\xi $, a distribution of the J-bonds would take
place, giving rise to the observed ``smoothing'' of the initial LAM
amplitude.

Second, we consider features ``B''. We assume that they result from a local
change in the exchange coupling ($J^{\prime }\neq J$). As studied by Eggert
(for $J^{\prime }<J$)\cite{11Eggert94}, LAM similar to the case $J^{\prime
}=0$ develop around such bond-defects. We suppose that the associated LAM
can be described by an expression similar to (1), but with a multiplying
factor $\alpha $ ($\alpha $ 
\mbox{$<$}%
1) to account for the expected smaller amplitude. As before, a parameter $%
\phi ^{\prime }$ describes a possible ``smoothing'' effect around the defect
position. Finally, we allow these defects to move along the chains. This
dynamical behavior is of crucial importance for the NMR lineshape as it
results in a ``flipping'' process of the hyperfine local field $%
h_{n}^{\prime }$ seen by the nuclear spins. Accordingly, the NMR signal of
each nuclear spin is composed of 1 or 2 distinct lines depending on the
flipping rate G with respect to the local hyperfine field $h_{n}^{\prime }$.
That ``motional narrowing'' is described as follows. For a nuclear spin at
site n in the chain, the NMR spectrum is written 
\[
I_{n}=\!{}\left[ D_{0}+K^{\prime }\left( \nu \right) \right] /\left\{ \left[
\nu +K^{\prime \prime }\left( \nu \right) \right] ^{2}+D_{0}^{2}+K^{\prime
}\left( \nu \right) ^{2}\right\} \text{ }\left( 2\right) 
\]
with $K^{\prime }\left( \nu \right) =%
\mathop{\rm Re}%
\left\{ K\left( \nu \right) \right\} $, $K^{\prime \prime }\left( \nu
\right) =%
\mathop{\rm Im}%
\left\{ K\left( \nu \right) \right\} $ and $\nu =\nu ^{0}-\gamma H/2\pi $,
where $\nu ^{0}$ is the experimental frequency and $D_{0}$ is the
``natural'' NMR width. $K(\nu )$ describes the effect of the dynamical local
field as $K(\nu )=(\gamma h_{n}^{\prime }/2\pi )^{2}k(\nu )$. Here, $k(\nu )$
is the Laplace representation of $k(t)$ which, as a function of time $t$,
describes the flipping process due to the motion of the defects. A ballistic
motion results in a linear exponential, $k(t)=exp(-t/G)$, while a diffusive
behavior gives $k(t)=exp(-\sqrt{t/G})$\cite{12Boucher82}. Such a single
moving defect in a long chain may create additional ``features'' in the
total NMR line, i.e., $I=\Sigma _{n}~I_{n}$. That model allows us to
reproduce very well the observed features ``B''. An example is given in Fig.
3b, where a comparison with the data obtained for $H$ // {\bf a}, at $\nu
^{0}=87~10^{-3}$ GHz and $T=30$ K is presented. The left side corresponds to
a diffusive behavior with the parameters: $\phi ^{\prime }=20$, $\alpha =0.6$%
, $G=2.8~10^{-4}$ GHz and $D_{0}=1.5~10^{-5}$ GHz. The ballistic model
(right side) gives usually a slightly poorer agreement (here with $\alpha
=0.45$, $G=1.5~10^{-4}$ GHz). That procedure applied to the different
experimental conditions yields an interesting result concerning $\phi
^{\prime }$. This parameter is independent on $H$, but it depends on $T$
according to $\phi ^{\prime }=(600\pm 100)/T\simeq \xi /2$. As for $\phi $,
this smoothing effect can be assumed to result from a local induced lattice
distortion around the bond-defects.

In presence of several such bond-defects, the above description applies as
long as the average distance between the defects remains large compared to $%
\xi $, i.e., a model of independent defects. At low $T$, however,
interactions between defects are expected: a magnetic interaction should
result from the overlap of the LAM structures and an elastic interaction
from the lattice distortion associated with the defects. As a balance
between these two interactions, we assume now that the defects form a
periodic structure. If it is the case, the shape of the resulting LAM will
depend on the number ${\it l}$ of spins between two neighboring
bond-defects. From the dashed line in Fig. 3a, approximate descriptions of
such a LAM are represented in Fig. 3c, for ${\it l}$ odd and even,
respectively. As expected, for ${\it l}$ even, a node occurs at the middle
position. For ${\it l}$ odd, the LAM amplitude remains finite at all
positions. In the former case, the resulting NMR line displays a single
narrow peak (Fig. 3d, right side). In the latter case, however, as the local
field seen by the nuclear spins never cancels, a double structure develops
at the centre of the NMR line (left side). For $H$ // {\bf a} , in
particular (see Fig. 1), such a central double structure is clearly observed
below $30$ K (and well reproduced by our model). For $H$ // {\bf c}, the NMR
line observed at $T=30$ K (Fig. 2a) is more complex. In addition to the
overlap of the three components of the quadrupolar splitting\cite
{2Takigawa97}, it can be analyzed as composed of two contributions as
illustrated in Fig. 2b. For one contribution (the dotted line), each
component displays a narrow central peak (similar to the case described in
Fig. 3d, right, for ${\it l}$ even) and for the other, it displays a central
double structure (as in Fig. 3d, left, for ${\it l}$ odd). At lower $T$,
however, only the double peak structure (features ``C'') remains (see Fig.
2c). The contribution from segments with ${\it l}$ even is seen to decrease
rapidly with $T$, and one is led to conclude that a periodic structure with $%
{\it l}$ odd becomes energetically preferable.

A complete description of the NMR line at low $T$ can now be proposed. A
chain of $N$ spins is considered, which contains a number $\eta $ of
bond-defects forming a periodic LAM structure with ${\it l}$ odd. The
dynamical behavior, i.e., the function $K(\nu )$ in Eq. 2, is now to be
considered as describing the dynamics (or the thermal fluctuations) of that
whole periodic LAM structure\cite{13Gaussian}. To account for the additional
static local fields induced by the chain edges, expressions for $h_{n}$ and $%
h_{N-n}$ [given by Eq. (1)] are added to the applied field $H$ in Eq. 2. The
total NMR line, $I=\Sigma _{n}$ $I_{n}$, is then calculated by varying the
parameters $\alpha $, $G$, ${\it l}$ and $D_{0}$. The parameters $\alpha $
and $G$ are used to adjust the width of feature ``B''. The shape of feature
``C'' is then determined by varying ${\it l}$ and $D_{0}$, keeping in mind
that the narrowing effect, i.e., the parameter $G$, plays here a crucial
role on the occurrence or not of such a small splitting at the centre of the
NMR line. Finally the number $\eta $ of bond-defects (with the condition $%
\eta {\it l}<N$) is used to adjust the intensity of features ``A''
relatively to that of the main line. With that procedure, one obtains the
open dots in Figs. 1 and 2. The agreement with the experiments is in general
excellent. Particular confidence in this approach is provided by the high
sensitivity of the final agreement to only a few parameters, essentially $%
{\it l}$ and $G$ (compare the different lines in Fig. 3d, left). At low
temperature, when features ``C'' are clearly visible, one always obtains the
following determination: $\eta {\it l}\simeq N$ ($\simeq 1800\pm 200$). This
means that the periodic LAM structure develops practically over all the
chain length. In that case, an evaluation of the density of the bond-defects
is simply given by $\rho \simeq {\it l}^{-1}$. For $H$ // {\bf a}, one
obtains a rather accurate determination of the parameters $\alpha $ and $G$.
They do not depend appreciably on $T$ and $H$: $\alpha =0.60\pm 0.04$, $%
G=(2.8\pm 0.2)$ $10^{-4}$ GHz. For $H$ // {\bf c}, the determination is much
more uncertain ($\alpha \simeq 0.46$ and $G\simeq 4.2$ $10^{-4}$ GHz) as a
small contribution with ${\it l}$ even is ignored in our description\cite
{14oddeven}. The parameter $D_{0}$ remains also constant ($D_{0}\simeq 1.5$ $%
10^{-5}$ GHz for $H$ // {\bf a}, and $D_{0}\simeq 5.6$ $10^{-5}$ GHz for $H$
// {\bf c}), except at the lowest temperature ($\simeq 20$ K) where larger
values (by a factor $\simeq 2$) provide a better agreement. At low $T$,
however, the description represented in Fig. 3c becomes insufficient when
the overlap of the LAM becomes very large. The main variation is finally
observed on the density of bond-defects $\rho \simeq l^{-1}$, which
decreases almost linearly with $T$. This result is well established for $H$
// {\bf a} (Fig. 2d). The values obtained for $H$ // {\bf c} are also shown
in that figure, though they are more uncertain for the reason explained above%
\cite{14oddeven}. A first question raised by the present description
concerns the source of the bond-defects ($J^{\prime }<J)$. Are they related
to the interstitial excess oxygen which characterizes that compound? There
are two reasons to rule out that possibility. First, at the lowest
temperature, $T=20$ K, the density of bond-defects ($\rho \simeq 2.8$ $%
10^{-3}$) remains more than ten times larger than the concentration of the
residual spin-1/2 impurities ($\simeq 1.3$ $10^{-4}$) after annealing.
Second, the observed T dependence of $\rho $ tells us that the number of
bond-defects is not a constant but it decreases with T. Remarkably, $\rho $
is observed to decrease linearly with T, as does the inverse of the spin
correlation length $\xi ^{-1}\simeq 2T/J$. That last remark strongly
supports our assumption that the bond-defects are an intrinsic property of
the ``real'' spin chains in Sr$_{2}$CuO$_{3}$.\ 

In the spin-Peierls phenomenology - and a priori above $T_{N}$, Sr$_{2}$CuO$%
_{3}$ is a SPS in its U phase - the description of the lattice is usually
considered in its linear approximation. The lattice modes are described by
simple phonon branches (characteristic frequency $\nu _{p}$) and the
spin-lattice coupling is considered in its adiabatic limit ($\nu _{p}<J$)%
\cite{4Bray83}. A more realistic description, however, should take into
account the non-linear properties of the lattice. They may give rise to
additional ``localized'' (and rather mobile) excitations (lattice solitons,
for instance). In the case of uniform chains, one kind of such non-linear
excitations are of particular interest: the so-called ``localized modes'' as
defined in \cite{15Sievers88}. They correspond to local oscillations of
atoms, at high frequency and with large amplitude (breather like). Their
presence in the lattice of an antiferromagnetic chain would result in
bond-defects as those discussed above\cite{16particular}. Alternatively, if
the SP phenomenology is considered in its non-adiabatic limit ($\nu _{p}>J$%
), non-linearities might be also expected from the spin-lattice coupling
itself. In that case, as discussed recently, the SP transition, instead of
being driven by a phonon softening, would be associated with a ``critical''
central peak\cite{17Gros98}, which usually signals the presence of
non-linear excitations. Well above $T_{sp}$, could this effective
non-linearity explain the presence of localized excitations? The present NMR
results may provide the first evidence of magneto-elastic localized modes in
UQHC. A recent numerical investigation of such bond-defects in UQHC\cite
{18Nishino} has already established the following results: i) the expected
magnetic interactions between two (static) bond-defects are attractive, ii)
a LAM\ structure with ${\it l}$ odd is energetically more favorable, iii)
the LAM amplitude of an {\it l}-even structure decreases rapidly at low
temperature. All these results agree with our analysis of the NMR line.
Together with the possible repulsive interactions due to the lattice, they
reinforce our proposition that a periodic LAM structure takes place at low $%
T $ in UQHC. At that point, it is worth reminding that a double-peak
structure, similar to our features ``C'', has been observed in the SPS CuGeO$%
_{3}$\cite{19Fagot98} well above $T_{sp}$, i.e., in its U phase. Our
conclusion is that the presence of a LAM structure is strongly supported by
the NMR results. We propose a model - bond-defects and a periodic LAM
structure - which explains both features ``B'' and ``C'' in an unique
consistent model.\ That model deserves certainly to be comforted
experimentally and theoretically. For instance, the quasi-static periodic AF
structure of the LAM could be observed by neutron diffraction. Any
description of such a ``real'' UQHC should consider explicitly the subtle
relations between the spins and the lattice.

One of us (JPB) thanks G. Uhrig and T. Ziman for illuminating discussions,
and S. Miyashita and the authors of Ref.\cite{18Nishino} for communicating
their results. Y. Ajiro is acknowledged for his invitation at Kyushu
University, and the Japan Society for the Promotion of Science for its
financial support.

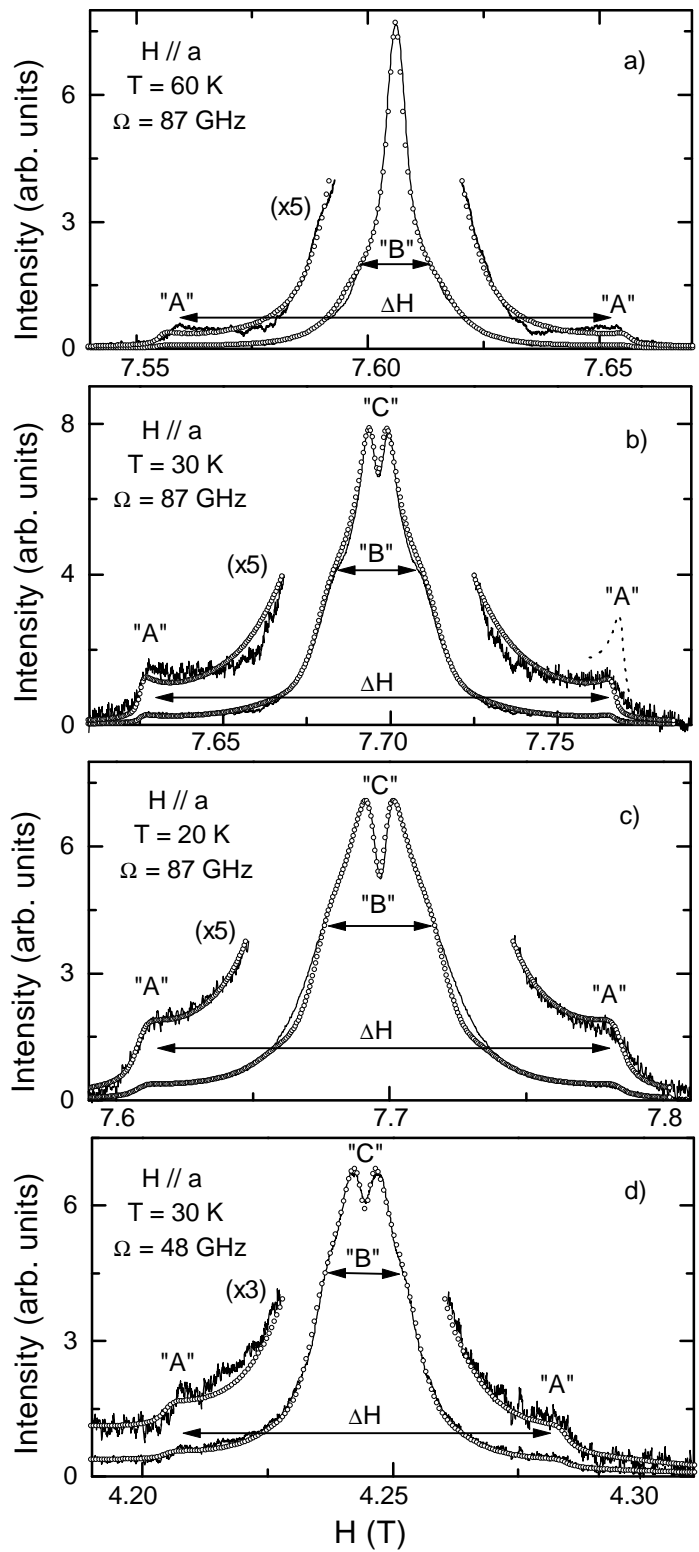
\begin{figure}[tbp]
\caption{NMR signals (solid lines) obtained at different frequencies $\nu
^{0},$for H // {\bf a} compared to calculated values (open dots). ``A'',
``B'' and ``C'' denote the ``features'' to be explained. In b), the dotted
line describing feature ``A'' corresponds to $\phi =0$, while the
calculation is obtained with $\phi =\xi $ (see dotted and solid lines in
Fig. 3a). }
\end{figure}

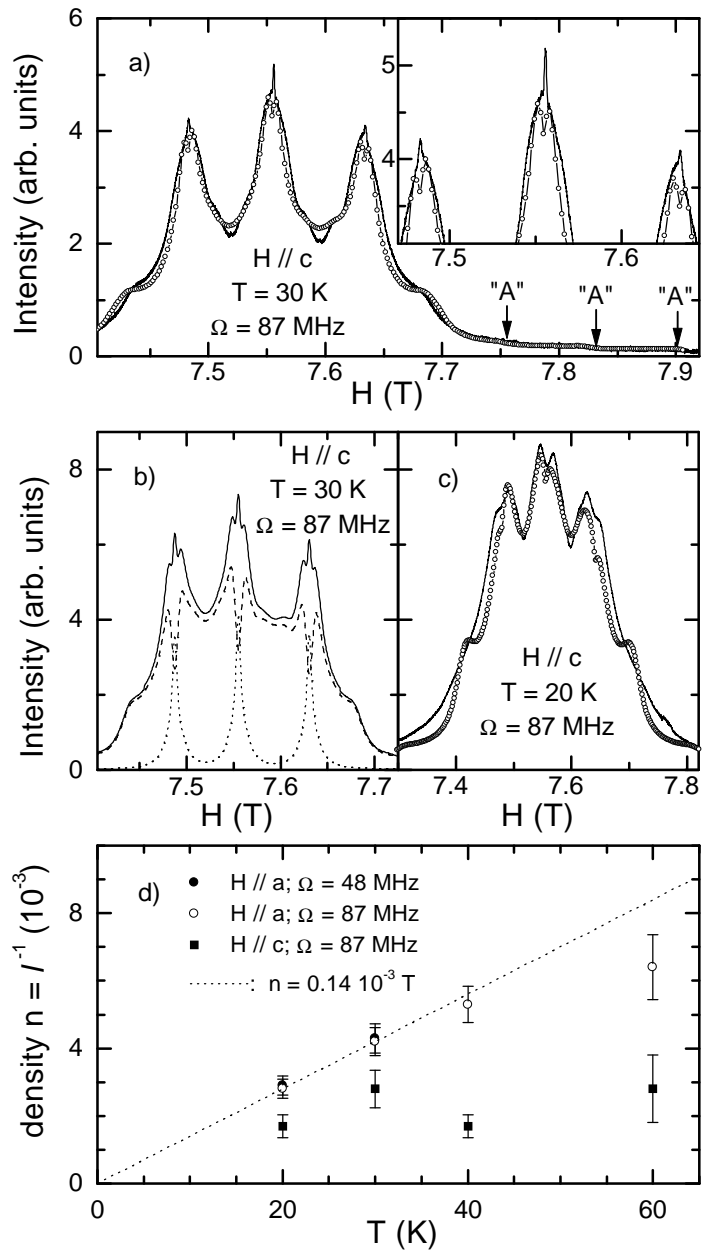
\begin{figure}[tbp]
\caption{a) NMR line for H // {\bf c}. The inset shows the details of the
central parts of the three hyperfine components. The open dots are
calculated for a periodic LAM with ${\it l}$ odd only. b) Example of a
spectrum resulting from LAM with both ${\it l}$ even (dotted) and odd
(dashed line), with relative intensity 0.13. c) Comparison between
calculation for ${\it l}$ odd only (open dots) and experiment (solid line)
obtained at $T=20$ K. d) $T$ dependence of the defect density $\rho $ for a
model of a periodic LAM structure with ${\it l}$ odd. The dotted line is a
linear fit with the data for H // {\bf a}. The data for H // {\bf c} are
also shown (see text).}
\end{figure}

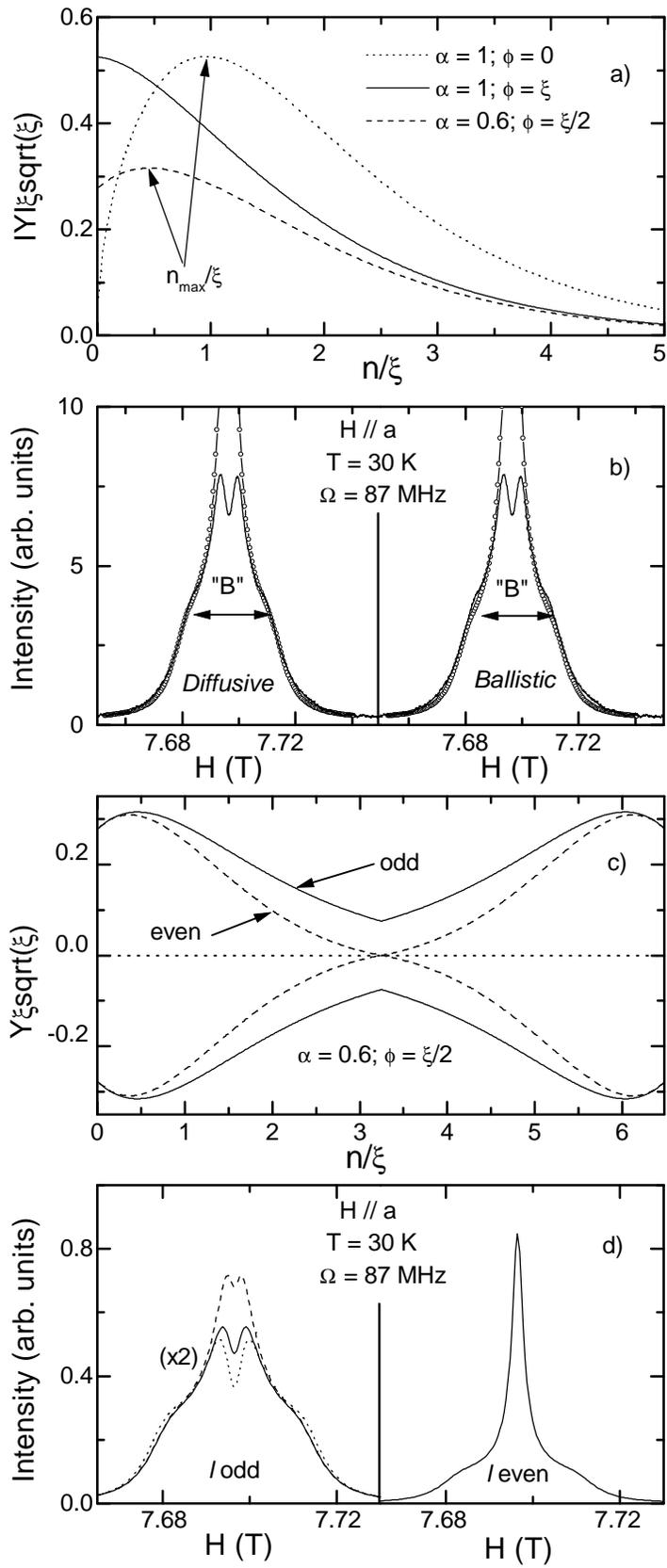
\begin{figure}[tbp]
\caption{a) Contour of the local field (in absolute value) as a function of
the site number n apart from a bond defect: $Y=h_{n}^{\prime }/(BA_{\beta
}H) $ and $\xi $, $\alpha $ and $\phi $ are defined in text. b) Description
of feature ``B'' by independent mobile bond-defects in the diffusive and
ballistic models (lines with open dots) compared to experimental data (solid
line). c) Contour of the local field between bond-defects with ${\it l}$ odd
and even as used in our calculations. d) NMR lineshapes resulting from
periodic LAM with ${\it l}$ odd and even: the solid lines are calculated for
the experimental conditions of Fig. 1b, with $G=2.8$ $10^{-4}$ GHz for ${\it %
l}=241$ (left) and ${\it l}=240$ (right); the dotted ($G=2.2$ $10^{-4}$ GHz,
same ${\it l}=241$) and dashed (${\it l}=261$, same $G=2.810^{-4}$ GHz)
lines illustrate the high sensitivity of these parameters on the final lineshape.}
\end{figure}


\begin{references}
\bibitem{13Gaussian}  Gaussian fluctuations can also be considered.
Different values will be obtained for the parameters, but the general
behavior will not change qualitatively.
\bibitem{16particular}  In particular, we expect the exchange coupling to be
locally reduced and according to [11], a value $\alpha $ $\simeq $ 0.6
corresponds to J/J' $\simeq $ 0.5.
\end{references}
\end{document}